# A New Algebraic Method to Search Irreducible Polynomials Using Decimal Equivalents of Polynomials over Galois Field GF($p^q$)

Sankhanil Dey [1] and Ranjan Ghosh[2]
Institute of Radio Physics and Electronics
University of Calcutta, 92 A. P. C. Road, Kolkata – 700 009
[1] sankhanil12009@gmail.com, [2]rghosh47@yahoo.co.in

**Abstract.** Irreducible polynomials play an important role till now, in construction of 8-bit S-Boxes in ciphers. The 8-bit S-Box of Advanced Encryption Standard is a list of decimal equivalents of Multiplicative Inverses (MI) of all the elemental polynomials of a monic irreducible polynomial over Galois Field GF($2^8$) [1]. In this paper a new method to search monic Irreducible Polynomials (IPs) over Galois fields GF($p^q$) has been introduced. Here the decimal equivalents of each monic elemental polynomial (ep), two at a time, are split into the p-nary coefficients of each term, of those two monic elemental polynomials. From those coefficients the p-nary coefficients of the resultant monic basic polynomials (BP) have been obtained. The decimal equivalents of resultant basic polynomials with p-nary coefficients are treated as decimal equivalents of the monic reducible polynomials, since monic reducible polynomials must have two monic elemental polynomials as its factor. The decimal equivalents of polynomials belonging to the list of reducible polynomials are cancelled leaving behind the monic irreducible polynomials. A non-monic irreducible polynomial is computed by multiplying a monic irreducible polynomial by α where α ∈ GF($p^q$) and assumes values from 2 to (p-1).

## 1. Introduction:

A basic polynomial BP(x) over finite field or Galois Field GF($p^q$) is expressed as,

$$BP(x) = a_q x^q + a_{q-1} x^{q-1} + \text{- - -} + a_1 x + a_0.$$

B(x) has (q+1) terms, where $a_q$ is non-zero and is termed as the leading coefficient [2]. A polynomial is monic if $a_q$ is unity, else it is non-monic. The GF($p^q$) have ($p^q - p$) elemental polynomials ep(x) ranging from p to ($p^q$ -1) each of whose representation involves q terms with leading coefficient $a_{q-1}$. The expression of ep(x) is written as,

$$ep(x) = a_{q-1} x^{q-1} + \text{- - -} + a_1 x + a_0 \text{, where } a_1 \text{ to } a_{q-1} \text{ are not simultaneously zero.}$$

Many of BP(x), which has an elemental polynomial as a factor under GF($p^q$), are termed as reducible. Those of the BP(x) that have no factors are termed as irreducible polynomials IP(x) [3][4] and is expressed as,

$$IP(x) = a_q x^q + a_{q-1} x^{q-1} + \text{- - -} + a_1 x + a_0 \text{, where } a_q \neq 0.$$

In Galois field GF($p^q$), the decimal equivalents of the basic polynomials of extension q vary from $p^q$ to ($p^{q+1}$ - 1) while the elemental polynomials are those with decimal equivalents varying from p to ($p^q - 1$). Some of the monic basic polynomials are irreducible, since it has no monic elemental polynomial as a factor. The method in this paper is to look for the decimal equivalents of the reducible polynomials with multiplication, addition and modulus, of the p-nary coefficients of each term of each two monic elemental polynomials to obtain the decimal equivalent of the p-nary coefficients of each term of the resultant monic basic polynomial. The resultant monic basic polynomials are termed as reducible polynomials, since it is the product of two monic elemental polynomials. The polynomials belonging to the list of reducible polynomials are cancelled leaving behind the monic irreducible polynomials. A non-monic irreducible polynomial is computed by multiplying a monic irreducible polynomial by α where α ∈ GF(p) and assumes values from 2 to (p-1). In literatures, to the best knowledge of the present authors, there is no mention of a paper in which the composite polynomial method is translated into an algorithm and in turn into a computer program.

Since 1967 researchers took algorithmic initiatives, followed by computational time-complexity analysis, to factorize basic polynomials on GF($p^q$) with a view to get irreducible polynomials, many of them are probabilistic [5][6][7][8] in nature and few of them are deterministic [9][10]. One may note that the deterministic algorithms are able to find all irreducible polynomials, while the probabilistic ones are able to find many, but not all. The irreducible polynomial over GF($2^8$) was first used in cryptography for designing an invertible 8-bit S-Box of AES [11][12][13]. The technique involves finding all multiplicative inverses under an irreducible polynomial is available in [14][15][16][17].

For convenient understanding, the proposed algebraic method is presented in Sec. 2 for p=7 with q=7. The method can find all monic and non-monic irreducible polynomials IP(x) over GF($7^7$). Sec. 3 is demonstrates the obtained results to show that the proposed searching algorithm is actually able to search for any extension of the Galois field with any prime of GF($p^q$), where p= 3, 5, 7,....,101,..,p and q= 2, 3, 5, 7,…,101,….q. In Sec.4 and 5, the conclusion of the paper and the references are illustrated. A sample list of the decimal equivalents of the monic irreducible polynomials over Galois Field GF($7^7$) is given in Appendix.

## 2 Algebraic method to find Irreducible Polynomials over GF($p^q$)

The basic idea of the algebraic method is to split the decimal equivalents of each monic elemental polynomial of a monic Basic Polynomial, any two at a time, into p-nary coefficients of each term of those monic elemental polynomials. Then multiply the p-nary coefficients to obtain each coefficients of the term with equal degree of each term of monic basic polynomials. All the multiplied results are then added to obtain the decimal coefficients of each term of the resultant monic basic polynomial. The decimal coefficients of each term of resultant monic basic polynomial are then reduced to p-nary coefficients of each term of that polynomial. The decimal equivalents of resultant monic basic polynomials with p-nary coefficients of each term are the decimal equivalents of the monic reducible polynomials since it has two monic elemental polynomials as its factor. The polynomials belonging to the list of monic reducible polynomials are cancelled leaving behind the monic irreducible polynomials. A non-monic irreducible polynomial is computed by multiplying a monic irreducible polynomial by α where α $\in$ GF(p) and assumes values from 2 to (p-1). The algebraic method to find the monic Irreducible polynomial over GF($7^7$) is demonstrated in section 2.2 and the extension to any Galois Field GF($p^q$) is described in section 2.2. The general method to develop each block of the algorithm of the algebraic method is demonstrated in section 2.3. The pseudo code for the above algebraic method is described in section 2.4.

### 2.1 Algebraic method to find Irreducible Polynomials over Galois Field GF($7^7$).

Here the interest is to find the monic irreducible polynomials over Galois Field or GF($7^7$).where p=7 is the prime field and q=7 is the extension of that prime field. Since the indices of multiplicand and multiplier are added to obtain the product. The extension q=7 can be demonstrated as a sum of two integers $d_1$ and $d_2$, The degree of highest degree term present in elemental polynomials of GF($7^7$) is (q-1) = 6 to 1, since the polynomials with highest degree of term 0, are constant polynomials and they do not play any significant role here, so they are neglected. Hence the two set of monic elemental polynomials for which the multiplication is a monic basic polynomial where p=7, q=7, have the degree of highest degree terms $d_1$, $d_2$ where, $d_1$=1,2,3, and the corresponding values of $d_2$ are, 6,5,4. Number of coefficients in the monic basic polynomial BP = (q+1)=(7+1) =8; they are defined as $BP_0$, $BP_1$, $BP_2$, $BP_3$, $BP_4$, $BP_5$, $BP_6$, $BP_7$, the value of the suffix also indicates the degree of the term of the monic basic polynomial. For monic polynomials $BP_7$= 1. Total number of blocks is the number of integers in $d_1$ or $d_2$, i.e. 3 for this case.

Coefficients of each term in the $1^{st}$ monic elemental polynomial $EP^0$ where, $d_1=1$; are defined as $EP_0^0$, $EP_1^0$. Coefficients of each term in the $2^{nd}$ monic elemental polynomial $EP^1$ where $d_2=6$; are defined as $EP_0^1$, $EP_1^1$, $EP_2^1$, $EP_3^1$, $EP_4^1$, $EP_5^1$, $EP_6^1$. The value in suffix also gives the degree of the term of the monic elemental polynomials.
Now, the algebraic method is as follows,

**$1^{st}$ block:**

$$BP_0 = (EP_0^0 * EP_0^1) \%7.$$
$$BP_1 = (EP_0^0 * EP_1^1 + EP_1^0 * EP_0^1) \%7.$$
$$BP_2 = (EP_0^0 * EP_2^1 + EP_1^0 * EP_1^1) \%7.$$
$$BP_3 = (EP_0^0 * EP_3^1 + EP_1^0 * EP_2^1) \%7.$$
$$BP_4 = (EP_0^0 * EP_4^1 + EP_1^0 * EP_3^1) \%7.$$
$$BP_5 = (EP_0^0 * EP_5^1 + EP_1^0 * EP_4^1) \%7.$$
$$BP_6 = (EP_0^0 * EP_6^1 + EP_1^0 * EP_5^1) \%7.$$
$$BP_7 = (EP_1^0 * EP_6^1) \%7 = 1.$$

Now the given basic monic polynomial is,
$$BP(x) = BP_7 x^7 + BP_6 x^6 + BP_5 x^5 + BP_4 x^4 + BP_3 x^3 + BP_2 x^2 + BP_1 x^1 + BP_0 x^0.$$
$$Decm\_eqv(BP(x)) = BP_7 * 7^7 + BP_6 * 7^6 + BP_5 * 7^5 + BP_4 * 7^4 + BP_3 * 7^3 + BP_2 * 7^2 + BP_1 * 7^1 + BP_0 * 7^0.$$

Coefficients of each term in the $1^{st}$ elemental monic polynomial $EP^0$ where, $d_1=2$; are defined as $EP_0^0$, $EP_1^0$, $EP_2^0$. Coefficients of each term in the $2^{nd}$ elemental monic polynomial $EP^1$ where $d_2=5$; are defined as $EP_0^1$, $EP_1^1$, $EP_2^1$, $EP_3^1$, $EP_4^1$, $EP_5^1$. The value in suffix also gives the degree of the term of the monic elemental polynomials.
Now, the algebraic method is as follows,

**$2^{nd}$ block:**

$$BP_0 = (EP_0^0 * EP_0^1) \%7.$$
$$BP_1 = (EP_0^0 * EP_1^1 + EP_1^0 * EP_0^1) \%7.$$
$$BP_2 = (EP_0^0 * EP_2^1 + EP_1^0 * EP_1^1 + EP_2^0 * EP_0^1) \%7.$$
$$BP_3 = (EP_0^0 * EP_3^1 + EP_1^0 * EP_2^1 + EP_2^0 * EP_1^1) \%7.$$
$$BP_4 = (EP_0^0 * EP_4^1 + EP_1^0 * EP_3^1 + EP_2^0 * EP_2^1) \%7.$$
$$BP_5 = (EP_0^0 * EP_5^1 + EP_1^0 * EP_4^1 + EP_2^0 * EP_3^1) \%7.$$
$$BP_6 = (EP_1^0 * EP_5^1 + EP_2^0 * EP_4^1) \%7.$$
$$BP_7 = (EP_2^0 * EP_5^1) \%7 = 1.$$

Now the given basic monic polynomial is,
$$BP(x) = BP_7 x^7 + BP_6 x^6 + BP_5 x^5 + BP_4 x^4 + BP_3 x^3 + BP_2 x^2 + BP_1 x^1 + BP_0 x^0.$$
$$Decm\_eqv(BP(x)) = BP_7 * 7^7 + BP_6 * 7^6 + BP_5 * 7^5 + BP_4 * 7^4 + BP_3 * 7^3 + BP_2 * 7^2 + BP_1 * 7^1 + BP_0 * 7^0.$$

Coefficients of each term in the $1^{st}$ monic elemental polynomial $EP^0$ where, $d_1=3$; are defined as $EP_0^0$, $EP_1^0$, $EP_2^0$, $EP_3^0$. Coefficients of each term in the $2^{nd}$ monic elemental polynomial $EP^1$ where $d_2=4$; are defined as $EP_0^1$, $EP_1^1$, $EP_2^1$, $EP_3^1$, $EP_4^1$. The value in suffix also gives the degree of the term of the monic elemental polynomials.
Now, the algebraic method is as follows,

**$3^{rd}$ block:**

$$BP_0 = (EP_0^0 * EP_0^1) \%7.$$
$$BP_1 = (EP_0^0 * EP_1^1 + EP_1^0 * EP_0^1) \%7.$$
$$BP_2 = (EP_0^0 * EP_2^1 + EP_1^0 * EP_1^1 + EP_2^0 * EP_0^1) \%7.$$
$$BP_3 = (EP_0^0 * EP_3^1 + EP_1^0 * EP_2^1 + EP_2^0 * EP_1^1 + EP_3^0 * EP_0^1) \%7.$$
$$BP_4 = (EP_0^0 * EP_4^1 + EP_1^0 * EP_3^1 + EP_3^0 * EP_1^1 + EP_2^0 * EP_2^1) \%7.$$
$$BP_5 = (EP_1^0 * EP_4^1 + EP_2^0 * EP_3^1 + EP_3^0 * EP_2^1) \%7.$$
$$BP_6 = (EP_2^0 * EP_4^1 + EP_3^0 * EP_3^1) \%7.$$
$$BP_7 = (EP_3^0 * EP_4^1) \%7 = 1.$$

Now the given basic monic polynomial is,

$$BP(x) = BP_7 x^7 + BP_6 x^6 + BP_5 x^5 + BP_4 x^4 + BP_3 x^3 + BP_2 x^2 + BP_1 x^1 + BP_0 x^0$$
$$\text{Decm\_eqv}(BP(x)) = BP_7 * 7^7 + BP_6 * 7^6 + BP_5 * 7^5 + BP_4 * 7^4 + BP_3 * 7^3 + BP_2 * 7^2 + BP_1 * 7^1 + BP_0 * 7^0$$

In this way the decimal equivalents of all the monic basic polynomials or monic reducible polynomials are pointed out. The polynomials belonging to the list of reducible polynomials are cancelled leaving behind the irreducible polynomials. Non-monic irreducible polynomials are computed by multiplying a monic irreducible polynomial by $\alpha$ where $\alpha \in GF(p)$ and assumes values from 2 to 6.

## 2.2 General Algebraic method to find Irreducible Polynomials over Galois Field $GF(p^q)$.

Here the interest is to find the monic irreducible polynomials over Galois Field $GF(p^q)$, where p is the prime field and q is the extension of the field. Since the indices of multiplicand and multiplier are added to obtain the product. The extension q can be demonstrated as a sum of two integers, $d_1$ and $d_2$, The degree of highest degree term present in elemental polynomials of $GF(p^q)$ is (q-1) to 1, since the polynomials with highest degree of term 0, are constant polynomials and they do not play any significant role here, so they are neglected. Hence the two set of monic elemental polynomials for which the multiplication is a monic basic polynomial, have the degree of highest degree terms $d_1$, $d_2$ where, $d_1 = 1,2,3,...,(q/2-1)$, and the corresponding values of $d_2$ are, (q-1), (q-2), (q-3),...,q-(q/2-1). Number of coefficients in the monic basic polynomial BP = (q+1); they are defined as $BP_0$, $BP_1$, $BP_2$, $BP_3$, $BP_4$, $BP_5$, $BP_6$, $BP_7$........, $BP_q$, the value of the suffix also indicates the degree of the term of the monic basic polynomial. For monic polynomials $BP_q = 1$.

Coefficients of each term in the 1st monic elemental polynomial $EP^0$, where, $d_1 = 1,2,.....,(q/2-1)$; are defined as $EP_0^0$, $EP_1^0$,......., $EP_{(q/2-1)}^0$. Coefficients of each term in the 2nd monic elemental polynomial $EP^1$ where $d_2 = $ (q-1), (q-2), (q-3),...,q-(q/2-1); are defined as $EP_0^1$, $EP_1^1$, $EP_2^1$, $EP_3^1$, $EP_4^1$, ... , $EP_{q-(q/2-1)}^1$. The value in suffix also gives the degree of the term of the monic elemental polynomials. Total number of blocks is the number of integers in $d_1$ or $d_2$, i.e. (q/2-1) for this example.

Now, the algebraic method for $(q/2-1)^{th}$ block is as follows,

**$(q/2-1)^{th}$ block:**
$$BP_0 = (EP_0^0 * EP_0^1) \%p.$$
$$BP_1 = (EP_0^0 * EP_1^1 + EP_1^0 * EP_0^1) \%p.$$
$$BP_2 = (EP_0^0 * EP_2^1 + EP_1^0 * EP_1^1 + EP_2^0 * EP_0^1) \%p.$$
$$BP_3 = (EP_0^0 * EP_3^1 + EP_1^0 * EP_2^1 + EP_2^0 * EP_1^1 + EP_3^0 * EP_0^1) \%p.$$
..................................................................
..................................................................
$$BP_{q-1} = (EP_0^0 * EP_{(q-1)}^1 + EP_1^0 * EP_{(q-2)}^1 +............+ EP_{(q/2-1)}^0 * EP_{(q-1)-(q/2-1)}^1) \%p.$$
$$BP_q = (EP_{(q/2-1)}^0 * EP_{q-(q/2-1)}^1) \%p.$$

Now the given basic monic polynomial is,
$$BP(x) = BP_q x^q + BP_{q-1} x^{q-1} +....+ BP_5 x^5 + BP_4 x^4 + BP_3 x^3 + BP_2 x^2 + BP_1 x^1 + BP_0 x^0.$$
$$\text{Decm\_eqv}(BP(x)) = BP_q * p^q + BP_{q-1} * p^{q-1} +....+ BP_5 * p^5 + BP_4 * p^4 + BP_3 * p^3 + BP_2 * p^2 + BP_1 * p^1 + BP_0 * p^0.$$

Similarly all the decimal equivalents of all the resultant basic polynomials or reducible polynomials for all a and its corresponding b values are calculated. The polynomials belonging to the list of reducible polynomials are cancelled leaving behind the irreducible polynomials. Non-monic irreducible polynomials are computed by multiplying a monic irreducible polynomial by $\alpha$ where $\alpha \in GF(p)$ and assumes values from 2 to (p-1).

**2.3 General Method to develop each block of the Algorithm of the New Algebraic Method.**

Prime field: p
Extension of the field: q.
$d_1 = 1,2,3,...........,(q/2-1)$.
$d_2 = $ (q-1), (q-2), (q-3).,...,q-(q/2-1).

Number of terms in 1st elemental polynomial: $N(d_1)$.
Number of terms in 1st elemental polynomial: $N(d_2)$.
Number of terms in Basic Polynomial: p.
Coefficients of Basic polynomial= $BP_{indx}$, where $1 < indx < p$
Coefficients of Elemental polynomials = $EP_{indx\_i}$, where $1 < i < 2$.

**Here,**

$$N(d_1) = N(d_2) = \text{Total number of blocks.}$$

Each coefficient of basic polynomial can be derived as follows,

$$BP_{indx} = (\Sigma EP_{indx\_1}{}^{p1} + EP_{indx\_2}{}^{p2}) \% p \dots\dots\dots\dots\dots\dots\dots\dots\dots\dots\dots\dots\dots\dots\dots\dots(i)$$

where$1 < indx < p$, $1 < indx1 < (q/2-1)$, $(q-1) > indx2 > q-(q/2-1)$.
$0 < p1 < N(d1)-1$, $0 < p2 < N(d2)-1$, and $indx = indx1+indx2$.

### 2.4. Pseudo code:

The pseudo code of the $(q/2-1)^{th}$ block of above algebraic code for Galois Field $GF(p^q)$ is described as follows, where ep[0] and ep[1] are the arrays of all possible decimal equivalents of 1st and 2nd monic elemental polynomials respectively. $EP^0$, $EP^1$ are the arrays consists of P-nary coefficients of 1st and 2nd monic elemental polynomials respectively. BP is the array consists of P-nary coefficients of the resultant monic basic polynomial. Decm_eqv(BP(x)) is the decimal equivalent of the resultant monic basic polynomial.

```
for(ep[0]=p..p^(q/2-1),ep[1]=p^(q-1)...p^(q-(q/2-1));ep[0]<2*p..p^(q/2-1),ep[0]<2*p^(q-1)...p^(q-(q/2-1));ep[0]++,ep[1]++){
  for(indx[0]=ep[0];indx[0]<2*ep[0];indx[0]++){
    coeff_conv_1st_deg (indx[0],EP^0);
    for(indx[1]=ep[1];indx[1]<2*ep[1];indx[1]++){
      coeff_conv_2nd_deg (indx[1], EP^1);
          BP_0= (EP_0^0 * EP_0^1) %p;
          BP_1= (EP_0^0 * EP_1^1+ EP_1^0 * EP_0^1) %p;
          BP_2= (EP_0^0 * EP_2^1+ EP_1^0 * EP_1^1+ EP_2^0 * EP_0^1) %p;
          BP_3= (EP_0^0 * EP_3^1+ EP_1^0 * EP_2^1+ EP_2^0 * EP_1^1+ EP_3^0 * EP_0^1) %p;
          ...........................................................
          ...........................................................
          BP_{q-1}= (EP_0^0 * EP_{q-1}^1+ EP_1^0 * EP_{q-2}^1+.........+ EP_{(q/2-1)}^0 * EP_{(q-1)-(q/2-1)}^1) %p;
          BP_q= (EP_{(q/2-1)}^0 * EP_{q-(q/2-1)}^1) %p;
BP(x) = BP_q x^q+ BP_{q-1} x^{q-1}+....+ BP_5 x^5+ BP_4 x^4+ BP_3 x^3+ BP_2 x^2+ BP_1 x^1+ BP_0 x^0;
Decm_eqv(BP(x))= BP_q *p^q+ BP_{q-1} *p^{q-1}+....+ BP_5* p^5+ BP_4 *p^4+ BP_3 *p^3+ BP_2 *p^2+ BP_1 *p^1+ BP_0 *p^0;
indx[2]++;
      End for.
    End for.
  End for
```

### 3. Results.

The algebraic method or the above pseudo code has been tested on $GF(7^3)$, $GF(11^3)$, $GF(101^3)$ and $GF(7^5)$. Numbers of monic Irreducible polynomials are same as in hands on as well as previous calculations [18]. The list of Numbers of monic irreducible polynomials are given below for the above four Extended Galois Fields. The list of all Irreducible monic basic polynomials of four extended Galois fields are available in reference [19][20][21][22]. The Sample list of monic Irreducible Polynomial over $GF(7^7)$ is given in Appendix and also available in the link given[23].

| GF | $GF(7^3)$ | $GF(11^3)$ | $GF(101^3)$ | $GF(7^5)$ |
|---|---|---|---|---|
| **Number of IP** | 112 | 440 | 343400 | 5712 |

## 4. Conclusion.

To the best knowledge of the present authors, there is no mention of a paper in which the composite polynomial method is translated into an algorithm and turn into a computer program. The new algebraic method is a much simpler method similar to composite polynomial method to find monic irreducible polynomials over Galois Field $GF(p^q)$. It is able to determine decimal equivalents of the monic irreducible polynomials over Galois Field with a larger value of prime, also with large extensions. So this method can reduce the complexity to find monic Irreducible Polynomials over Galois Field with large value of prime and also with large extensions of the prime field. So this would help the crypto community to build S-Boxes or ciphers using irreducible polynomials over Galois Fields with a large value of prime, also with the large extensions of the prime field.

# Appendi

A sample list of Decimal Equivalents of Monic Irreducible Polynomials over Galois Field $GF(7^7)$ is given below.

```
823586  823587  823588  823589  823590  823591  823595  823596  823598  823601  823602
823604  823607  823611  823612  823614  823618  823619  823622  823623  823625  823630
823631  823633  823635  823636  823638  823643  823644  823646  823649  823653  823654
823658  823659  823661  823665  823666  823668  823670  823674  823675  823678  823679
823681  823684  823685  823687  823691  823692  823696  823698  823700  823701  823706
823708  823709  823713  823715  823716  823719  823721  823722  823726  823727  823731
823735  823737  823738  823740  823744  823745  823749  823750  823752  823755  823756
823758  823762  823763  823765  823770  823771  823773  823775  823779  823780  823782
823783  823785  823790  823792  823793  823796  823797  823801  823803  823805  823806
823810  823812  823813  823817  823818  823822  823825  823827  823828  823833  823835
823836  823838  823840  823841  823846  823848  823849  823852  823853  823857  823859
823860  823864  823867  823869  823870  823873  823875  823876  823882  823884  823885
823887  823892  823894  823899  823902  823905  823908  823913  823916  823919  823923
823926  823930  823931  823932  823933  823937  823940  823946  823948  823950  823954
823959  823962  823964  823965  823966  823967  823973  823975  823978  823980  823988
823990  823994  823997  824001  824003  824007  824010  824013  824017  824020  824021
824022  824023  824027  824029  824034  824037  824042  824045  824050  824051  824052
824053  824055  824057  824063  824065  824070  824074  824079  824081  824085  824088
824091  824094  824097  824098  824099  824100  824107  824109  824113  824115  824118
824122  824125  824127  824132  824134  824139  824142  824147  824149  824154  824157
824161  824165  824169  824170  824171  824172  824177  824179  824182  824185  824188
824190  824196  824200  824202  824205  824211  824212  824213  824214  824217  824219
824226  824228  824231  824234  824238  824241  824246  824247  824252  824255  824260
824261  824267  824268  824272  824274  824275  824277  824279  824283  824290  824291
824295  824298  824302  824303  824308  824309  824314  824315  824317  824319  824322
824326  824332  824333  824337  824338  824342  824343  824345  824347  824349  824353
824358  824361  824364  824365  824371  824375  824379  824380  824385  824389  824394
824395  824398  824399  824405  824407  824409  824410  824412  824415  824419  824423
824428  824429  824433  824437  824441  824442  824451  824452  824454  824455  824457
824459  824463  824466  824469  824473  824475  824476  824484  824485  824489  824491
824493  824494  824497  824501  824503  824506  824513  824514  824517  824521  824525
824529  824531  824532  824538  824541  824547  824548  824555  824556  824559  824561
824563  824564  824566  824570  824573  824578  824580  824585  824589  824590  824594
824599  824603  824604  824610  824611  824615  824616  824619  824620  824625  824627
824633  824634  824637  824638  824643  824644  824646  824647  824650  824655  824659
824661  824667  824668  824675  824676  824678  824679  824681  824682  824687  824689
824695  824697  824700  824701  824706  824711  824716  824717  824721  824722  824724
824725  824727  824729  824734  824739  824741  824742  824749  824751  824758  824759
824763  824764  824769  824770  824772  824773  824779  824781  824783  824788  824794
824795  824799  824801  824805  824806  824814  824815  824818  824819  824826  824827
824829  824830  824833  824835  824839  824841  824849  824850  824853  824858  824861
824862  824867  824869  824874  824875  824884  824885  824889  824890  824892  824893
824895  824900  824903  824905  824910  824911  824918  824919  824925  824926  824930
824935  824939  824940  824944  824949  824951  824956  824958  824959  824962  824963
824967  824969  824972  824977  824979  824980  824982  824983  824987  824988  824997
```

824998 825003 825005 825010 825011 825014 825019 825022 825023 825031 825033
825037 825039 825042 825043 825045 825046 825053 825054 825059 825060 825066
825067 825071 825073 825077 825078 825084 825089 825091 825093 825099 825100
825102 825103 825106 825107 825113 825114 825121 825123 825130 825131 825133
825138 825143 825145 825147 825148 825150 825151 825157 825158 825161 825166
825171 825172 825175 825177 825183 825185 825190 825191 825193 825194 825196
825197 825204 825205 825211 825213 825217 825222 825225 825226 825228 825229
825234 825235 825238 825239 825245 825247 825253 825254 825261 825262 825268
825269 825274 825277 825282 825283 825288 825291 825295 825298 825301 825303
825304 825306 825308 825309 825311 825313 825316 825317 825324 825325 825331
825334 825340 825341 825343 825347 825351 825355 825358 825359 825366 825369
825371 825375 825378 825379 825381 825383 825387 825388 825396 825397 825400
825404 825406 825409 825413 825415 825417 825418 825420 825421 825430 825431
825435 825439 825443 825444 825448 825452 825457 825460 825462 825463 825465
825467 825473 825474 825477 825478 825483 825487 825492 825493 825498 825502
825507 825508 825511 825514 825519 825523 825525 825527 825529 825530 825534
825535 825539 825540 825546 825550 825553 825555 825557 825558 825563 825564
825569 825570 825574 825577 825581 825582 825589 825593 825595 825599 825603
825606 825610 825613 825616 825621 825624 825627 825630 825635 825637 825642
825645 825646 825647 825648 825653 825655 825658 825659 825660 825661 825667
825670 825672 825676 825682 825684 825687 825690 825693 825695 825700 825701
825702 825703 825707 825711 825715 825718 825723 825725 825730 825733 825738
825740 825742 825744 825750 825754 825757 825759 825763 825765 825772 825773
825774 825775 825778 825781 825784 825787 825794 825796 825798 825802 825807
825809 825815 825817 825819 825820 825821 825822 825827 825830 825835 825838
825840 825842 825849 825850 825851 825852 825855 825859 825862 825865 825869
825871 825875 825878 825882 825884 825892 825894 825897 825899 825905 825906
825907 825908 825910 825913 825918 825922 825924 825926 825932 825935 825941
825943 825945 825946 825948 825954 825956 825957 825975 825977 825978 825980
825981 825983 825987 825988 825990 825994 825995 826001 826006 826011 826012
826018 826019 826022 826027 826029 826030 826037 826038 826039 826041 826043
826046 826052 826053 826058 826062 826065 826069 826073 826074 826078 826081
826085 826086 826087 826089 826094 826097 826100 826104 826108 826109 826110
826115 826116 826117 826121 826125 826129 826132 826134 826135 826138 826139
826142 826144 826149 826152 826155 826157 826162 826164 826170 826173 826177
826179 826183 826186 826187 826188 826194 826195 826200 826201 826204 826206
826209 826211 826213 826216 826221 826222 826229 826230 826233 826234 826235
826236 826241 826243 826246 826248 826254 826257 826258 826261 826264 826265
826267 826269 826276 826278 826281 826283 826284 826286 826290 826293 826298
826300 826302 826307 826310 826313 826319 826321 826323 826324 826325 826327
826330 826332 826337 826338 826347 826348 826353 826354 826355 826356 826358
826359 826366 826367 826372 826374 826377 826379 826381 826383 826388 826390
826395 826397 826401 826402 826405 826407 826410 826411 826415 826417 826421
826426 826428 826429 826431 826433 826442 826444 826445 826449 826450 826452
826458 826459 826461 826463 826467 826468 826471 826473 826474 826484 826485
826486 826488 826492 826493 826499 826501 826506 826507 826515 826516 826519
826521 826530 826531 826533 826536 826542 826545 826548 826549 826555 826558
826561 826564 826565 826566 826569 826573 826577 826578 826582 826585 826589
826590 826593 826594 826596 826600 826605 826608 826611 826612 826613 826620

```
826621  826622  826625  826628  826635  826636  826638  826642  826646  826649  826654
826655  826659  826663  826667  826668  826669  826671  826674  826678  826682  826685
826689  826692  826697  826698  826699  826701  826702  826703  826708  826711  826716
826719  826722  826723  826725  826726  826730  826731  826732  826734  826739  826741
826743  826746  826753  826755  826757  826759  826765  826769  826773  826775  826778
826779  826786  826787  826788  826789  826793  826794  826803  826804  826806  826809
826811  826813  826815  826817  826822  826823  826828  826830  826834  826836  826841
826843  826845  826846  826849  826851  826858  826860  826863  826866  826867  826870
826871  826874  826883  826884  826888  826890  826891  826893  826897  826901  826902
826905  826906  826908  826911  826913  826914  826925  826926  826936  826937  826942
826944  826948  826949  826953  826954  826955  826957  826963  826964  826967  826972
826974  826977  826978  826979  826982  826983  826988  826993  826997  826998  827003
827004  827011  827013  827019  827020  827026  827027  827031  827035  827037  827041
827044  827045  827051  827055  827058  827061  827062  827063  827067  827070  827074
827075  827076  827079  827083  827086  827089  827094  827095  827097  827098  827101
827105  827107  827110  827117  827118  827119  827121  827123  827129  827130  827131
827133  827135  827140  827142  827146  827151  827153  827157  827159  827164  827167
827172  827173  827177  827178  827185  827187  827189  827191  827193  827195  827202
827203  827209  827210  827212  827213  827214  827215  827220  827221  827224  827226
827227  827230  827235  827237  827243  827245  827247  827249  827250  827252  827255
827257  827261  827266  827268  827269  827273  827277  827278  827280  827283  827284
827286  827289  827290  827292  827297  827299  827300  827319  827321  827325  827326
827333  827334  827338  827343  827346  827347  827352  827355  827356  827357  827362
827363  827366  827369  827370  827371  827373  827377  827380  827381  827387  827391
827395  827399  827404  827405  827410  827413  827415  827418  827423  827427  827430
827431  827433  827434  827436  827439  827443  827447  827452  827453  827454  827460
827461  827462  827465  827467  827473  827475  827478  827482  827485  827490  827493
827494  827495  827497  827499  827501  827507  827510  827517  827518  827524  827525
827527  827529  827531  827535  827537  827539  827541  827542  827550  827551  827555
827556  827557  827558  827563  827565  827569  827571  827572  827574  827579  827581
827585  827587  827590  827591  827594  827598  827599  827602  827604  827609  827619
827621  827622  827625  827626  827628  827633  827634  827636  827641  827642  827644
827646  827647  827651  827661  827662  827663  827665  827667  827671  827676  827677
827682  827685  827688  827692  827699  827700  827703  827707  827710  827713  827716
827719  827723  827724  827725  827733  827734  827735  827739  827742  827746  827749
827751  827752  827754  827755  827759  827763  827765  827767  827775  827777  827779
827782  827789  827791  827794  827795  827796  827798  827802  827804  827807  827809
827811  827814  827817  827819  827825  827826  827829  827830  827836  827837  827838
827839  827842  827843  827851  827852  827857  827860  827861  827866  827868  827871
827873  827877  827879  827881  827882  827884  827886  827892  827894  827899  827900
827903  827905  827907  827908  827913  827914  827916  827919  827923  827924  827926
827927  827929  827933  827934  827938  827957  827958  827961  827962  827963  827965
827970  827971  827978  827980  827986  827987  827989  827990  827996  828001  828003
828004  828005  828007  828013  828015  828018  828021  828024  828029  828034  828036
828040  828043  828045  828047  828052  828054  828057  828060  828061  828066  828067
828075  828076  828077  828078  828083  828084  828087  828088  828094  828096  828098
828101  828106  828109  828111  828115  828118  828119  828123  828124  828127  828131
828133  828138  828140  828143  828144  828146  828148  828151  828153  828154  828157
828161  828162  828166  828167  828169  828171  828172  828174  828178  828180  828181
```

```
828199  828201  828209  828210  828214  828215  828221  828223  828228  828229  828234
828235  828236  828238  828245  828246  828249  828253  828255  828258  828259  828260
828263  828266  828270  828271  828278  828281  828283  828286  828292  828293  828300
828301  828302  828305  828306  828307  828313  828316  828318  828322  828325  828326
828329  828330  828332  828335  828340  828343  828353  828354  828356  828362  828364
828365  828369  828371  828372  828374  828375  828377  828388  828389  828391  828396
828400  828402  828405  828411  828412  828418  828419  828423  828426  828431  828435
828437  828438  828439  828441  828444  828446  828452  828454  828459  828462  828466
828469  828473  828475  828479  828481  828486  828489  828490  828491  828494  828497
828498  828502  828504  828507  828509  828514  828516  828523  828525  828529  828532
828533  828535  828537  828538  828540  828545  828546  828549  828550  828556  828561
828563  828568  828570  828571  828580  828581  828585  828586  828587  828589  828593
828594  828595  828600  828603  828606  828610  828613  828617  828621  828624  828628
828629  828630  828633  828634  828637  828638  828640  828642  828645  828651  828652
828657  828658  828664  828665  828672  828673  828675  828677  828680  828683  828684
828685  828686  828690  828693  828698  828701  828706  828708  828713  828715  828717
828718  828719  828721  828724  828729  828731  828733  828738  828741  828742  828747
828749  828753  828755  828761  828763  828766  828771  828774  828775  828778  828780
828781  828783  828785  828788  828789  828794  828795  828796  828798  828804  828805
828809  828810  828815  828817  828823  828825  828833  828834  828838  828841  828843
828846  828851  828852  828853  828857  828858  828861  828862  828867  828868  828869
828871  828875  828879  828882  828885  828886  828892  828893  828900  828901  828909
828910  828913  828915  828918  828922  828923  828924  828925  828927  828929  828931
828936  828937  828939  828948  828952  828953  828955  828957  828958  828963  828965
828966  828969  828970  828972  828984  828987  828990  828993  828997  829000  829001
829002  829006  829009  829012  829016  829019  829020  829027  829028  829034  829035
829043  829044  829046  829050  829053  829057  829061  829062  829063  829065  829068
829071  829075  829079  829084  829086  829089  829090  829091  829093  829095  829097
829105  829107  829110  829114  829116  829119  829125  829127  829131  829133  829138
829140  829147  829149  829151  829153  829159  829162  829163  829165  829167  829169
829170  829173  829174  829177  829181  829182  829186  829187  829193  829194  829195
829197  829204  829205  829210  829211  829217  829219  829221  829226  829228  829229
829230  829237  829240  829242  829245  829251  829254  829257  829260  829266  829267
829268  829270  829271  829273  829274  829281  829282  829284  829285  829291  829294
829296  829299  829300  829301  829302  829305  829307  829309  829313  829314  829321
829322  829326  829330  829331  829341  829342  829344  829347  829348  829352  829354
829356  829357  829361  829362  829364  829377  829378  829382  829385  829386  829387
829390  829391  829397  829398  829405  829407  829410  829415  829420  829421  829425
829426  829428  829429  829433  829434  829435  829439  829443  829445  829449  829452
829455  829459  829462  829469  829470  829471  829473  829475  829477  829482  829483
829491  829492  829494  829495  829505  829506  829509  829511  829513  829515  829516
829517  829518  829522  829523  829525  829529  829530  829534  829537  829539  829540
829545  829546  829548  829558  829559  829561  829571  829572  829581  829582  829585
829589  829593  829597  829599  829602  829603  829604  829606  829610  829615  829618
829620  829624  829627  829629  829636  829638  829642  829643  829644  829646  829649
829651  829655  829660  829663  829666  829672  829674  829677  829678  829681  829683
829685  829686  829688  829690  829691  829694  829698  829700  829706  829708  829711
829716  829718  829723  829727  829729  829733  829734  829740  829741  829746  829749
829750  829751  829755  829756  829763  829764  829767  829770  829774  829777  829781
```

```
829785  829789  829793  829797  829798  829799  829803  829804  829806  829807  829812
829813  829814  829817  829819  829821  829827  829828  829830  829831  829841  829842
829846  829847  829851  829853  829855  829858  829859  829860  829861  829866  829867
829869  829881  829882  829884  829887  829889  829890  829893  829894  829898  829900
829901  829903  829914  829918  829921  829924  829925  829926  829929  829933  829935
829939  829945  829946  829949  829950  829958  829961  829963  829968  829971  829973
829978  829979  829980  829982  829986  829988  829991  829993  829998  830002  830006
830009  830014  830016  830020  830022  830026  830027  830030  830033  830035  830036
830038  830041  830042  830045  830050  830052  830054  830059  830062  830065  830069
830070  830071  830073  830075  830079  830082  830086  830093  830094  830098  830099
830104  830108  830110  830113  830118  830122  830127  830129  830131  830133  830139
830140  830141  830143  830148  830150  830154  830156  830159  830161  830163  830164
830167  830170  830171  830173  830175  830183  830185  830188  830190  830195  830197
830202  830203  830206  830211  830213  830218  830219  830226  830227  830229  830230
830231  830233  830236  830237  830245  830246  830250  830255  830260  830261  830262
830265  830268  830273  830276  830278  830281  830287  830290  830292  830293  830294
830299  830300  830302  830303  830307  830308  830313  830315  830317  830321  830322
830323  830324  830327  830330  830332  830334  830335  830345  830346  830350  830351
830355  830356  830358  830362  830364  830365  830369  830370  830374  830377  830378
830380  830390  830394  830395  830404  830409  830411  830412  830413  830415  830421
830423  830428  830430  830434  830437  830440  830443  830446  830448  830454  830455
830458  830460  830465  830469  830471  830475  830477  830483  830485  830488  830491
830492  830495  830496  830498  830500  830503  830505  830509  830511  830517  830518
830526  830527  830530  830531  830532  830534  830538  830539  830548  830549  830551
830555  830561  830562  830563  830565  830566  830569  830570  830573  830574  830575
830579  830582  830588  830591  830594  830597  830602  830603  830604  830605  830607
830609  830612  830617  830618  830622  830623  830628  830629  830635  830636  830642
830644  830646  830649  830650  830652  830657  830659  830660  830663  830665  830666
830670  830674  830675  830686  830687  830689  830699  830700  830706  830710  830714
830717  830719  830722  830723  830724  830726  830729  830734  830737  830742  830743
830747  830748  830750  830763  830765  830766  830770  830772  830773  830782  830783
830785  830791  830793  830794  830796  830800  830804  830805  830806  830810  830813
830817  830820  830825  830826  830827  830831  830835  830838  830839  830842  830843
830848  830850  830852  830853  830856  830860  830862  830867  830869  830873  830874
830877  830883  830885  830887  830889  830890  830892  830896  830897  830901  830905
830910  830913  830917  830920  830922  830926  830931  830932  830937  830939  830940
830941  830944  830945  830950  830953  830955  830957  830958  830964  830965  830971
830974  830976  830979  830980  830986  830987  830988  830989  830992  830997  831000
831001  831010  831011  831017  831018  831021  831022  831027  831032  831034  831036
831037  831038  831042  831045  831051  831053  831057  831059  831064  831066  831072
831074  831077  831080  831083  831084  831085  831088  831090  831092  831098  831101
831105  831107  831108  831109  831111  831114  831118  831123  831125  831127  831135
831137  831141  831142  831144  831146  831148  831149  831154  831155  831157  831169
831171  831172  831174  831175  831179  831188  831189  831192  831193  831195  831198
831202  831203  831204  831211  831214  831217  831221  831225  831226  831227  831231
831234  831238  831240  831245  831246  831249  831251  831256  831259  831261  831265
831268  831269  831274  831276  831279  831281  831283  831284  831286  831289  831294
831297  831300  831304  831309  831312  831317  831318  831321  831322  831323  831326
831330  831331  831336  831337  831346  831347  831349  831352  831354  831360  831361
```

```
831363  831364  831365  831366  831373  831374  831378  831380  831382  831387  831388
831394  831395  831401  831403  831406  831408  831409  831410  831414  831416  831420
831421  831426  831427  831434  831438  831442  831443  831447  831449  831450  831451
831454  831455  831462  831465  831469  831473  831475  831479  831483  831484  831485
831486  831489  831490  831497  831499  831501  831507  831508  831513  831514  831517
831519  831522  831526  831527  831531  831532  831540  831541  831546  831547  831556
831557  831559  831561  831567  831569  831570  831571  831573  831578  831580  831582
831587  831589  831594  831598  831601  831603  831604  831605  831610  831613  831618
831620  831623  831625  831629  831633  831634  831636  831637  831641  831645  831646
831648  831659  831661  831662  831665  831667  831668  831678  831681  831688  831689
831690  831693  831696  831699  831702  831706  831707  831708  831715  831718  831721
831722  831724  831725  831730  831732  831736  831738  831741  831742  831745  831750
831752  831755  831756  831758  831760  831762  831764  831769  831772  831773  831779
831780  831785  831786  831791  831793  831797  831798  831799  831802  831804  831809
831814  831815  831819  831820  831825  831827  831828  831829  831833  831837  831841
831843  831849  831851  831853  831858  831861  831863  831868  831871  831874  831876
831877  831883  831885  831886  831895  831899  831900  831903  831905  831906  831910
831911  831913  831924  831928  831930  831931  831933  831934  831939  831942  831945
831946  831947  831953  831956  831958  831962  831965  831966  831967  831973  831976
831977  831979  831981  831988  831990  831993  831996  831997  832001  832003  832007
832009  832010  832012  832014  832019  832021  832025  832032  832033  832035  832036
832037  832040  832043  832044  832050  832054  832057  832061  832063  832066  832074
832075  832078  832080  832082  832084  832085  832093  832094  832098  832100  832102
832105  832106  832114  832115  832116  832117  832122  832123  832126  832131  832133
832134  832135  832138  832141  832143  832149  832150  832157  832158  832162  832163
832169  832171  832175  832180  832185  832187  832191  832193  832197  832201  832203
832205  832206  832207  832211  832214  832218  832219  832221  832225  832227  832228
832231  832235  832236  832247  832249  832250  832252  832254  832255  832266  832270
832275  832278  832281  832282  832283  832289  832292  832294  832295  832297  832298
832302  832306  832308  832309  832310  832315  832317  832318  832320  832323  832325
832329  832332  832333  832338  832340  832343  832345  832351  832354  832355  832357
832362  832365  832369  832372  832376  832379  832380  832385  832386  832387  832390
832396  832397  832399  832403  832406  832409  832413  832414  832420  832422  832424
832429  832430  832434  832435  832442  832444  832446  832452  832453  832457  832458
832459  832460  832463  832467  832470  832473  832476  832477  832483  832485  832486
832487  832492  832493  832498  832502  832504  832508  832511  832513  832516  832521
832522  832529  832530  832533  832535  832537  832539  832540  832547  832548  832549
832550  832555  832556  832561  832563  832564  832565  832567  832569  832578  832579
832582  832583  832590  832591  832595  832596  832602  832607  832612  832614  832618
832621  832623  832625  832626  832627  832630  832634  832637  832639  832644  832646
832652  832654  832659  832661  832662  832667  832669  832670  832681  832682  832684
832686  832687  832691  832693  832697  832698  832709  832712  832714  832715  832716
832721  832724  832729  832732  832738  832739  832740  832742  832745  832750  832751
832753  832754  832756  832758  832763  832764  832766  832768  832772  832774  832777
832778  832781  832786  832788  832794  832796  832798  832801  832802  832805  832807
832812  832817  832819  832822  832827  832829  832830  832831  832834  832837  832840
832842  832850  832852  832854  832855  832859  832863  832865  832866  832876  832877
832879  832882  832884  832885  832891  832892  832894  832905  832906  832907  832911
832915  832919  832922  832924  832925  832926  832931  832934  832938  832939  832942
```

```
832943  832946  832949  832954  832956  832959  832962  832963  832967  832969  832973
832978  832981  832982  832985  832988  832990  832994  832996  832998  832999  833001
833002  833003  833006  833011  833012  833017  833020  833022  833026  833030  833033
833036  833039  833045  833046  833053  833054  833057  833058  833059  833060  833068
833069  833071  833074  833076  833082  833083  833086  833087  833093  833095  833097
833100  833101  833108  833110  833114  833116  833117  833118  833123  833125  833130
833131  833137  833138  833141  833142  833150  833152  833153  833162  833163  833165
833169  833170  833172  833185  833187  833188  833190  833191  833193  833198  833201
833204  833206  833212  833214  833219  833221  833225  833227  833233  833236  833239
833242  833243  833244  833246  833251  833256  833257  833260  833261  833267  833268
833277  833278  833281  833286  833289  833290  833291  833293  833298  833299  833302
833304  833307  833313  833314  833320  833321  833323  833325  833328  833333  833334
833338  833339  833340  833341  833346  833347  833352  833356  833358  833361  833365
833368  833373  833377  833381  833382  833386  833387  833388  833390  833393  833395
833401  833404  833405  833409  833411  833416  833418  833422  833425  833426  833428
833430  833435  833437  833438  833440  833443  833447  833451  833452  833453  833458
833461  833465  833468  833472  833473  833474  833478  833482  833484  833485  833488
833489  833494  833496  833499  833502  833505  833506  833507  833509  833514  833517
833519  833524  833529  833531  833533  833535  833541  833542  833548  833549  833554
833556  833561  833562  833563  833565  833569  833571  833578  833579  833586  833587
833592  833593  833596  833597  833603  833605  833608  833610  833611  833619  833620
833621  833622  833625  833626  833631  833633  833635  833640  833643  833646  833649
833653  833657  833659  833662  833667  833668  833673  833676  833677  833678  833682
833683  833689  833691  833694  833697  833698  833701  833706  833710  833712  833716
833717  833720  833723  833725  833729  833730  833732  833734  833736  833737  833740
833741  833745  833748  833753  833754  833755  833757  833760  833764  833768  833772
833773  833774  833779  833782  833785  833787  833788  833794  833795  833797  833800
833802  833803  833813  833814  833816  833820  833824  833825  833834  833838  833843
833844  833849  833850  833851  833853  833859  833860  833863  833866  833869  833873
833877  833881  833883  833885  833891  833893  833898  833901  833902  833905  833907
833911  833913  833915  833916  833921  833923  833926  833927  833930  833934  833937
833939  833940  833941  833947  833950  833955  833958  833963  833964  833965  833967
833970  833974  833975  833977  833978  833981  833982  833986  833988  833992  833993
833995  833997  833998  834009  834010  834012  834017  834018  834020  834033  834035
834040  834042  834045  834049  834052  834053  834054  834056  834058  834061  834065
834067  834074  834076  834083  834084  834088  834089  834096  834097  834100  834101
834110  834112  834114  834115  834116  834118  834121  834126  834129  834130  834131
834132  834139  834140  834142  834144  834146  834149  834150  834157  834158  834163
834166  834168  834172  834173  834178  834179  834186  834187  834193  834195  834198
834201  834202  834203  834205  834210  834213  834214  834222  834223  834226  834227
834233  834235  834237  834244  834245  834249  834250  834255  834257  834259  834265
834266  834268  834269  834270  834271  834276  834280  834282  834283  834289  834292
834293  834294  834299  834300  834303  834307  834310  834314  834319  834322  834324
834325  834328  834334  834336  834339  834341  834346  834347  834350  834354  834356
834359  834361  834362  834364  834366  834371  834373  834377  834381  834382  834384
834385  834387  834390  834396  834397  834398  834401  834404  834409  834413  834418
834419  834420  834424  834425  834427  834429  834430  834432  834443  834444  834448
834451  834452  834454  834458  834460  834461  834472  834473  834474  834476  834478
834482  834486  834488  834492  834494  834499  834504  834508  834510  834514  834517
```

```
834521  834522  834527  834532  834534  834537  834538  834539  834543  834545  834550
834551  834556  834557  834565  834566  834573  834574  834577  834579  834581  834585
834586  834594  834595  834597  834599  834601  834604  834605  834611  834612  834613
834614  834618  834622  834625  834629  834635  834636  834639  834642  834643  834644
834646  834647  834654  834658  834662  834665  834667  834669  834670  834672  834676
834678  834682  834683  834686  834689  834691  834698  834700  834702  834703  834706
834709  834714  834717  834721  834723  834726  834732  834733  834734  834737  834740
834745  834746  834748  834749  834751  834755  834761  834762  834763  834766  834768
834769  834773  834774  834776  834779  834780  834784  834793  834794  834796  834802
834803  834805  834816  834818  834821  834826  834828  834830  834836  834838  834842
834846  834850  834851  834852  834854  834857  834860  834863  834867  834871  834874
834881  834882  834885  834886  834887  834889  834893  834894  834898  834902  834906
834910  834915  834917  834919  834921  834923  834924  834927  834929  834934  834937
834938  834941  834943  834947  834949  834955  834956  834959  834961  834964  834971
834972  834973  834977  834980  834983  834986  834989  834990  834991  834998  835001
835003  835004  835006  835007  835011  835012  835014  835017  835018  835020  835031
835033  835034  835038  835042  835043  835045  835046  835050  835059  835061  835066
835069  835074  835075  835076  835078  835081  835085  835090  835092  835097  835099
835103  835105  835108  835109  835110  835112  835118  835120  835122  835123  835132
835133  835138  835139  835147  835148  835150  835155  835157  835160  835162  835165
835166  835171  835172  835178  835180  835182  835189  835190  835193  835194  835195
835196  835201  835202  835209  835211  835213  835218  835222  835225  835227  835228
835229  835231  835235  835238  835244  835246  835248  835250  835258  835259  835263
835265  835269  835270  835271  835273  835276  835278  835284  835285  835291  835292
835301  835302  835305  835306  835313  835314  835315  835316  835318  835319  835325
835327  835330  835332  835333  835342  835343  835346  835348  835350  835353  835356
835357  835358  835361  835362  835367  835370  835375  835379  835382  835385  835390
835393  835397  835398  835403  835405  835410  835411  835414  835418  835420  835423
835428  835430  835433  835434  835439  835441  835444  835445  835447  835449  835452
835453  835454  835458  835462  835465  835468  835475  835476  835477  835481  835484
835486  835487  835490  835491  835494  835497  835500  835504  835505  835507  835508
835510  835522  835524  835525  835530  835531  835533  835535  835537  835538  835558
835560  835561  835563  835564  835566  835570  835571  835573  835579  835581  835582
835593  835595  835596  835598  835601  835603  835605  835606  835613  835614  835620
835621  835626  835627  835633  835636  835638  835641  835642  835643  835644  835648
835649  835650  835654  835657  835661  835665  835668  835672  835675  835678  835683
835684  835685  835689  835690  835693  835694  835697  835698  835707  835708  835710
835715  835717  835722  835728  835729  835732  835733  835738  835740  835741  835742
835745  835746  835749  835753  835755  835762  835764  835769  835771  835774  835776
835780  835781  835784  835787  835789  835790  835792  835797  835799  835803  835805
835809  835812  835816  835819  835824  835826  835832  835834  835836  835837  835838
835841  835843  835847  835852  835855  835859  835860  835866  835867  835873  835876
835878  835882  835886  835888  835889  835890  835893  835896  835899  835902  835906
835908  835913  835915  835921  835923  835924  835925  835927  835932  835937  835939
835942  835945  835950  835953  835955  835956  835957  835960  835962  835965  835969
835973  835979  835980  835985  835986  835990  835992  835993  836004  836005  836009
836013  836014  836016  836019  836020  836022  836027  836029  836030  836043  836044
836050  836051  836056  836057  836061  836062  836067  836070  836072  836074  836075
836076  836077  836082  836084  836086  836088  836091  836097  836100  836104  836105
```

```
836106  836109  836110  836113  836114  836116  836117  836118  836124  836128  836131
836134  836140  836141  836145  836147  836148  836149  836152  836153  836161  836162
836168  836170  836174  836176  836179  836180  836187  836188  836191  836194
```